\documentclass[aps,pra,reprint,superscriptaddress,noeprint]{revtex4-2}

\usepackage[english]{babel}

\usepackage{amsmath}
\usepackage{amssymb}
\usepackage{wasysym}
\usepackage{graphicx}
\usepackage{hyperref}
\usepackage{xcolor}
\usepackage{physics}
\usepackage{units}
\usepackage[normalem]{ulem}

\usepackage[capitalise]{cleveref}  

\begin{document}

\title{Long-term stable laser injection locking for quasi-CW applications}

\author{Florian Kiesel}
\affiliation{Physikalisches Institut and Center for Integrated Quantum Science and Technology, Eberhard Karls Universit\"at T\"ubingen, 72076 Tübingen, Germany}

\author{Kirill Karpov}
\affiliation{Physikalisches Institut and Center for Integrated Quantum Science and Technology, Eberhard Karls Universit\"at T\"ubingen, 72076 Tübingen, Germany}

\author{Alexandre de Martino}
\affiliation{Physikalisches Institut and Center for Integrated Quantum Science and Technology, Eberhard Karls Universit\"at T\"ubingen, 72076 Tübingen, Germany}

\author{Jonas Auch}
\affiliation{Physikalisches Institut and Center for Integrated Quantum Science and Technology, Eberhard Karls Universit\"at T\"ubingen, 72076 Tübingen, Germany}

\author{Christian Gross}
\email[]{christian.gross@uni-tuebingen.de}
\affiliation{Physikalisches Institut and Center for Integrated Quantum Science and Technology, Eberhard Karls Universit\"at T\"ubingen, 72076 Tübingen, Germany}

\date{\today}

\begin{abstract}
	Generating high output powers while achieving narrow line single mode lasing are often mutual exclusive properties of commercial laser diodes.
	For this reason, efficient and scalable amplification of narrow line laser light is still a major driving point in modern laser system designs.
	Commonly, injection locking of high-power semiconductor laser diodes are used for this purpose.
	However, for many laser diodes it is very challenging to achieve stable operation
	of the injection locked state due to a complex interplay of non-linearities and thermal effects.
	Different approaches of active or passive stabilization usually require a large overhead of optical and electrical equipment and are not generally applicable.
	In our work we present a passive stabilization scheme, that is generally applicable, technically easy to implement and extremely cost-effective.
	It is based on the externally synchronized automatic acquisition of the optimal injection state.
	Central to our simple but powerful scheme is the management of thermalization effects during lock acquisition.
	By periodical relocking, spectrally pure amplified light is maintained in a quasi-CW manner over long timescales.
	We characterize the performance of our method for laser diodes amplifying $\unit[671]{nm}$ light and demonstrate the general applicability by confirming the method to work also for laser diodes at $\unit[401]{nm}$, $\unit[461]{nm}$ and $\unit[689]{nm}$.
	Our scheme enables the scaled operation of injection locks, even in cascaded setups, for the distributed amplification of single frequency laser light.
\end{abstract}

\maketitle

Applications in atomic, molecular, and optical (AMO) physics often require several narrow line laser beams  with output powers in the $\unit[100]{mW}$ range.
While tapered amplifiers offer a scalable solution in the red and near-infrared spectral range, they are not available at shorter wavelengths.
Narrow-line laser amplification becomes increasingly important in this spectral range, due to the rising popularity of alkaline-earth and alkaline-earth-like atoms for AMO / trapped ions experiments and quantum technologies \cite{takasu2003, takamoto2005optical, ludlow2006srstudy, KraftCalciumBEC2009, stellmer2009,aikawa2012bose, luDysBEC2015, cooper18, norcia19,  wilson22, haffner08}.
Injection locking of high-power laser diodes is a commonly used and cost-effective technique for this purpose.\\
The term injection locking describes the external forcing of the slave laser to the frequency of the  seed laser by optically injecting the seed into the slave \cite{liu2020optical}.
Successful injection locking requires the slave laser's parameters (e.g. temperature and current) to be set precisely to match the slave laser's cavity to the seed light.
Since the first successful demonstration for a laser in 1966 \cite{stover1966locking} various schemes of active and passive stabilisation have been developed.
To provide feedback to the system, they make use of many different properties such as polarization purity \cite{niederriter2021}, Fabry-P\'erot signals \cite{saxberg2016, shimasaki2019injection}, atomic vapor cell spectroscopy signals \cite{pagett_injection_2016}, interference filter transmission \cite{schkolnik2020, chen2021active}, intensity noise \cite{Chiow07} and beam shape \cite{Mishra23} of the injection locked laser beam.

All of these methods rely on a non-negligible amount of overhead equipment and at least partly depend on the specific diode which hinders scalability and general applicability. \\
We present a highly reliable and cost-efficient injection locking technique, that allows for long-term stability and is generally applicable to a large variability of slave diodes.
Key to the technique is its simplicity.
It only requires a microcontroller to read the slave laser diode's internal photo diode and to control the set point of the slave laser current.
The locking technique relies on the characteristic dependence of the photo diode signal as a function of the slave laser diode's current.
Using our robust method to assure thermalization, the injection locked state can be maintained long-term by periodically optimizing the current of the slave diode.
This process needs less than $\unit[0.5]{s}$ and can be synchronized to an external experimental cycle via a trigger.
We successfully verified our technique for laser diodes at $\unit[401]{nm}$, $\unit[461]{nm}$, $\unit[671]{nm}$ and $\unit[689]{nm}$ providing up to $\unit[190]{mW}$ and in serial operation of two slave diodes.
The general applicability and minimalistic hardware demand allows for a scalable operation of injection locks, enabling a distributed laser system for multi-beam power delivery.\\

\begin{figure*}[t]
	\centering
	\includegraphics{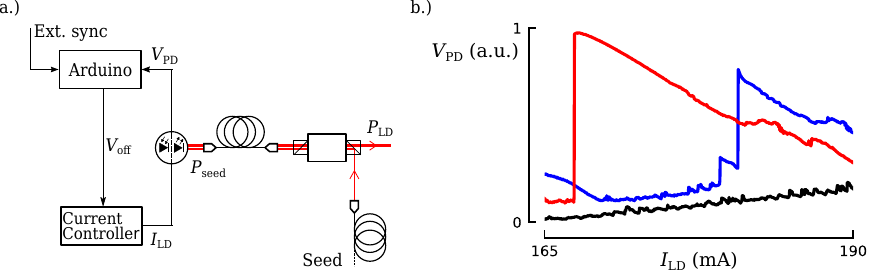}
	\caption{
		a.) Sketch of the system's optical and electronic setup.
		The optical mode of the seed light is overlapped with the mode of the LD on a FI and then mode matched with an optical fiber.
		The seed with power $P_{\text{seed}}$ is then injected into the LD.
		The amplified light travels through the overlapped path again and is then transmitted by the FI.
		An Arduino monitors the LD's internal PD signal ($V_{\text{PD}}$) and controls the LD's current $I_{\text{LD}}$ via $V_{\text{off}}$ on the current controller's modulation input. \\
		b.) Typical PD signal for the sweep of the LD current $I_{\text{LD}}$ up (blue) and down (red) with the presence of the seed light and without (black).
		With seed light present the PD signal ($V_{\text{PD}}$) shows the typical injection resonance features with a dominant jump on the low current side.
		A strong hysteresis of the system can well be observed.
		The sweep down injection resonance is at lower current, of greater width and features a small plateau region next to the jump point.
		The unperturbed LD signal shows no features but a constant slope due to higher output powers.
	}
	\label{fig:setup_sketch}
\end{figure*}

\subsection*{Experimental setup}

For the characterization of our injection locking method we focus on a slave laser diode (LD) at $\unit[671]{nm}$ (\textit{Ushio HL67001DG}).
We use the generated laser light for the laser cooling of lithium, of which the performance and stability serves as an independent check for the spectral quality of the generated light.
We operate six slave LDs in two stage serial configuration with output powers between $\unit[100]{mW}$ and $\unit[190]{mW}$.
The experimental setup for our injection locking scheme in reflection configuration~\cite{liu2020optical} is shown in Fig. \ref{fig:setup_sketch}.
A Faraday isolator (FI) is used to guide the seed light into the slave diode.
A single-mode polarization-maintaining optical fiber between the LD and FI assures quantifiable mode matching and allows for spatial separation of slave and seed.
This configuration poses no issues in the red, but leads to disturbing optical feedback for blue diodes.
In this case we instead placed the fiber at the seed port's transmission direction after the FI.
Also with this configuration mode overlap can be measured using the small amount of slave light reflected out of this FI port.
We operate at slave laser fiber coupling efficiencies of about  $\unit[50]{\%}$ and with seed light power of $\unit[2]{}$-$\unit[4]{mW}$ after the fiber.
The slave LD's operation is monitored using the integrated photo diode (PD).
We convert the photo current to a voltage using a standard transimpedance-amplifier circuit.
This voltage is monitored by a microcontroller (\textit{Arduino Due}).
The microcontroller controls the slave LD current $I_{\text{LD}}$ via the current controller's external modulation input.\\

\begin{figure*}[t]
	\centering
	\includegraphics{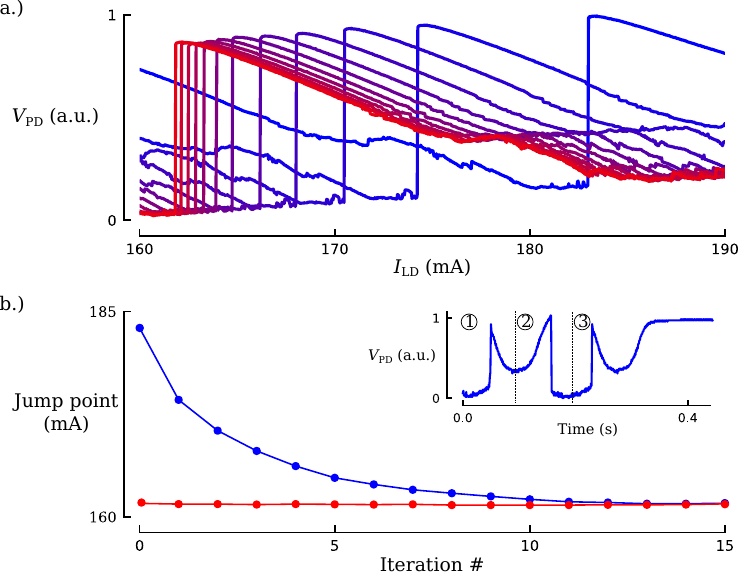}
	\caption{
		a.) Multiple sweep up PD signals for  a continuous application of the ILA of a previously unlocked system.
		The ILA was repeatedly started after a hold time of $\unit[0.5]{s}$.
		The traces are colored from blue to red for increasing iteration numbers.
		The convergence of the jump points are presented in b.).
		The blue curve displays the jump points of a.).
		It highlights the thermalization behavior in contrast to a system, which was unlocked for a few seconds only (red).
		After about 10 iterations the previously unlocked system has thermalized.
		The inset shows a typical PD signal of an ILA run.
		The encircled numbers of its individual steps are displayed at the top.
	}\label{fig:jump_points}
\end{figure*}

To measure the characteristic slave LD response in presence of seed light we sweep the LD current while monitoring the internal PD signal with the microcontroller.
This signal's linear dependence on $I_{\text{LD}}$ is distinctively modified with the presence of the seed light (see Fig.~\ref{fig:setup_sketch}).
The observed light level increases on the internal PD (decreases on an external PD), when the slave LD's cavity becomes resonant with the seed.
Injection locking forces the slave LD to emit at the frequency of the seed.
With increasing spectral purity the slave LD spectrum more and more equalizes with the one of the seed~\cite{hosoya2015injection}.\\
The pronounced asymmetric shape of the signal is explained by cavity expansion due to Joule heating and heating from resonant seed light to the cavity which also leads to hysteresis~\cite{saxberg2016}.
For sweeps from high to low current the jump in the PD signal occurs at smaller currents than for sweeps in the opposite direction.
The peak with its characteristic jump on the low current side is much more prominent for the sweep down.
More stable and spectrally more pure injection can be realized when ramping the slave current with falling slope close to the abrupt jump.
Some diodes (as the one studied here) also show a small plateau just before the jump.
The resonance feature appears repeatedly with similar shape for different slave LD current, which defines distinct operation currents.
The resonance can also be tuned via the temperature of the slave LD.\\
The sharp jump at the lower end of the resonance feature can be conveniently used to identify the optimal operating current using the microcontroller.\\

\subsection*{Current optimization algorithm}

To automatically find the optimal current with maximum spectral purity we use the microcontroller to ramp the laser diode current upon an external trigger.
The PD signal is synchronously recorded by one of the microcontroller's 12-bit analogue input pins.
The injection lock optimization algorithm (ILA) proceeds in three steps:\\
\begin{enumerate}
	\item Sweep up $I_{\text{LD}}$ with the resonance jump in the middle of the ramp.
	\item Sweep down $I_{\text{LD}}$ over the same interval while recording the PD signal and take the numerical derivative to identify the jump point in the resonance feature.
	\item Sweep up again and then sweep down to the optimal current which is chosen with a diode and work point dependent offset to the previously detected jump point.
\end{enumerate}
For this algorithm to work the slave LD current has to be manually set to roughly center the jump point in the ramp (1. step).
To account for longer timescale thermalization effects (see below) this is best done by continuously modulating the slave LD current with the same triangular waveform as used in the algorithm and to adjust the set point accordingly.
Furthermore, the speed and range of the ramps has to be adapted to the slave LD.
For the diode under study here, we hold every current step in the discrete ramps for about $\unit[0.1]{ms}$.
Executing this algorithm takes about $\unit[0.4]{s}$, then the laser current is kept constant.
We trigger the automatic optimization of the slave LD current periodically and synchronized to our experimental sequence to stay close to optimal spectral purity.
The frequency of this periodic optimization depends on the environmental conditions and the specific application.
For typical experiments with ultracold gases it can be done in the cooling sequence when no laser light is needed such that the optimization does not affect the cycle time.
A typical run of the ILA is depicted in Fig. \ref{fig:jump_points}. \\

\begin{figure*}[t]
	\centering
	\includegraphics{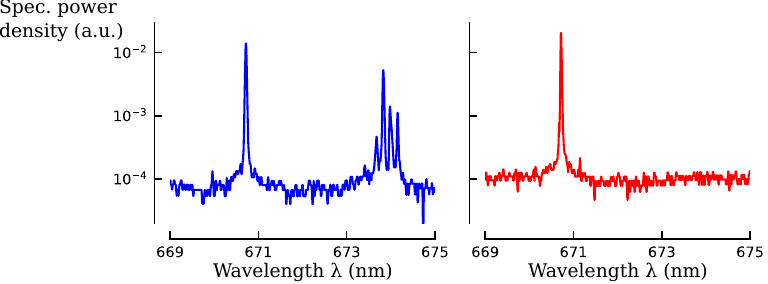}
	\caption{Spectral power density measurements with the LSA for a partially locked (blue) and fully locked (red) USHIO diode.
		For the partially locked case the spectral power is split up between the seed light mode at $\sim \unit[671]{nm}$ and the free spectrum of the USHIO between 673 nm and 675 nm.
		In this state, the powers in the seed mode and the free running modes are unstable.
		In the fully locked case the spectral power density of the free running USHIO vanishes within the noise limit.
		Here the LD lases on the seed light mode only.
	}\label{fig:lsa_traces}
\end{figure*}

\subsection*{Thermalization of the slave laser}

The position of the jump point is strongly dependent on the history of the system (Fig. \ref{fig:jump_points}).
Starting from an initially unlocked system, it moves to lower currents for continuous application of the triangular current ramps used in the optimization algorithm.
The location of the jump point converges to a constant value after about 10 iterations.
Thus, we implement a "cold start" option by holding the microcontroller trigger to run about 15 ramp cycles.
Afterwards the trigger is released and the algorithm finds the optimal current reliably.
When the system loses injection lock for only a short amount of time (tens of seconds) the jump point does not change significantly between consecutive iterations.\\

\subsection*{Characterization of the injection locking algorithm}

\begin{figure*}[t!]
	\centering
	\includegraphics{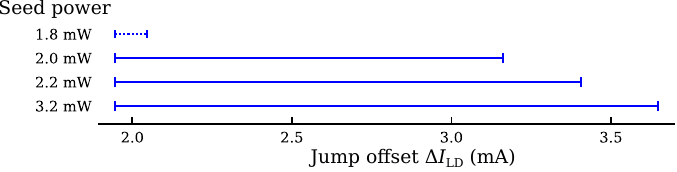}
	\caption{
		Spectral purity measurements for different seed powers and jump offset currents from the jump point.
		For $\unit[3.2]{mW}$, $\unit[2.2]{mW}$, $\unit[2.0]{mW}$ of seed powers the solid lines indicate the interval of jump offset currents, in which a fully injected system can be achieved.
		For all these three seed powers the lower current threshold is found to be $\unit[1.9]{mA}$.
		The high current threshold slightly decreases from $\unit[3.6]{mA}$, to $\unit[3.4]{mA}$ and $\unit[3.2]{mA}$ for seed power reduction.
		For jump currents higher than these thresholds the system is unstable.
		For a seed power of $\unit[1.8]{mW}$ the stable interval basically vanishes and a fully injected system can not be achieved anymore.
	}
	\label{fig:spec_pur_li}
\end{figure*}

We now investigate the performance of our automatic slave current optimization technique by the analysis of the optical spectrum for different parameters.
For measurement of the spectral purity we use on optical spectrum analyzer (\textit{HighFinesse Laser Spectrum Analyzer}).
The free running slave LD spectrum is about $\unit[2]{nm}$ wide around a center wavelength of $\unit[674]{nm}$ at a temperature of $\unit[22]{^\circ C}$.
It is well separated from the seed light at $\unit[671]{nm}$, see Fig.~\ref{fig:lsa_traces}.
To determine the spectral purity of the injection locked light, we compare the integrated spectral power densities $P_\mathrm{lock}$ and $P_\mathrm{free}$ of the injection locked and free running diode.
We define the spectral purity as $S = (P_\mathrm{lock} - P_\mathrm{free})/(P_\mathrm{lock} + P_\mathrm{free})$ and similarly the spectral impurity as $\bar{S} = 1-S$. \\
For $\bar{S}=0$ all the optical power is in the seed light's mode and we call the system fully injected locked.
On the contrary for $\bar{S}=1$ the slave LD is lasing on modes of its free spectrum exclusively. \\
We investigate the performance of the injection locking algorithm for different seed powers and offsets of the optimal current $I_{\text{LD}}$ from the detected jump point (see Fig.~\ref{fig:spec_pur_li}).
The presented data was taken for a current driver set point of $\unit[204]{mA}$, which provides $\unit[180]{mW}$ of optical power.
We assure thermalization of the slave LD as described above and measure the spectral purity after the algorithm finished.
For a seed power of $\unit[3.2]{mW}$ we find that reliable injection locking is obtained for offsets from the jump point larger than $\unit[1.9]{mA}$.
For smaller offsets the fully injection locked state is typically lost after a short amount of time.
For upt to $\unit[3.6]{mA}$ a fully injection locked system is obtained.
For larger offset currents the spectral purity decreases.\\
For reduced seed power the offset interval, in which the system is fully injection locked, decreases from the upper end.
Below $\unit[2]{mW}$, the fully injected state can no longer be realized.
The width of the stable region decreases non-linearly with seed power, a typical behavior for mode competition of a lasing system \cite{murakami2003cavity}.\\

\begin{figure*}[t!]
	\centering
	\includegraphics{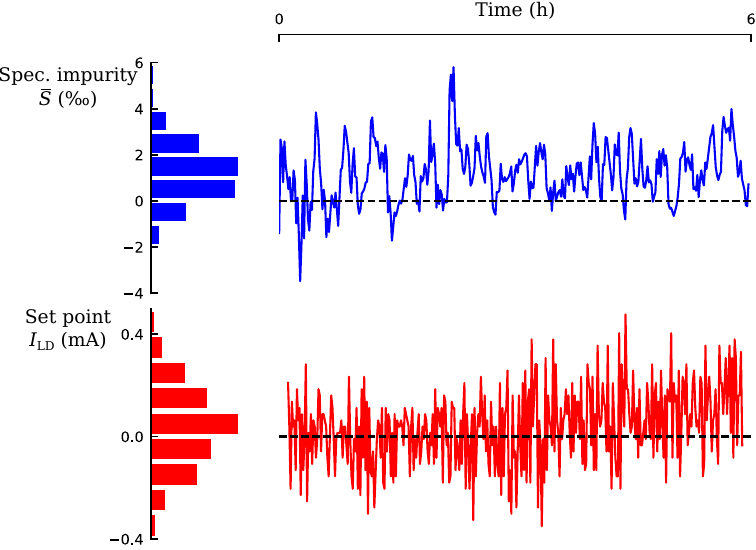}
	\caption{
		Long term stability measurement of a periodically ($T=\unit[1]{min}$) applied ILA for a total time of 6 hours.
		With a mean of $1.18\; \permil$ the spectral impurity never exceeds $6\; \permil$.
		This means, that the ILA never failed to bring the system to a fully injection locked state for 360 consecutive applications.
		For monitoring the jump current the mean of the first 20 recorded iterations was taken as a reference value.
		The set point stays within a $\pm \unit[0.4]{mA}$ range for the whole measurement.
		A drift of the system is observed after about 3 hours.
		This leads to an average of the setpoint for the last 20 iterations of  $\unit[0.18]{mA}$, while the spectral impurity is not affected.
		A histogram of the corresponding data is presented on the y-axis of the respective subplots.
	}\label{fig:long_term_spec}
\end{figure*}
To investigate the reliability of the algorithm under typical experimental conditions we set the seed power to $\unit[3.2]{mW}$ and the laser current driver to $\unit[204]{mA}$.
The offset current is chosen to be $\unit[2.7]{mA}$, well in the fully injection locked interval.
We periodically run our algorithm every minute to readjust the optimal current for a total time of six hours.
The spectral (im-)purity of the amplified light is recorded after each run of the algorithm (see Fig.~\ref{fig:long_term_spec}).
For the total number of 360 repetitions the algorithm never failed to bring the system to a fully injection locked state.
The measured impurity was consistently at the detection limit (below $6\; \permil$ with a mean of $1.18\; \permil$).
Detection noise explains the negative spectral impurity for some data points.
This measurement shows that our algorithm brings the system to a fully injection locked state with near unity probability.
Even in the unlikely case of a failure to reach full injection, the next periodically triggered optimization run will restore injection as long as the retriggering period is chosen shorter than the thermalization time of the diode.
We confirmed the spectral purity measurement with long-time monitoring of the performance of the entire laser cooling system.
The detected atom number after laser cooling is highly sensitive to the spectral purity, and we found it to be stable within detection noise of a few percent.\\

Our algorithm only fails without chance for recovery if the optimal current set point drifts outside the ramp interval which can be chosen to be several tens of mA wide.
This is unlikely, even at the scale of months, such that with our algorithm injection locks will not limit the stability of even complex laser systems enabling scalability and distributed amplification.
In our laser system for the laser cooling of lithium we tested this scalability by implementing two chained injection locks.
In this setup, the seed laser is amplified by one slave LD, which then seeds another slave LD.
We confirmed that by applying our algorithm sequentially on the two serially connected laser diodes, reliable injection locking of the entire system can still be assured.\\

We also tested our scheme for other slave laser diodes in the context of laser cooling of erbium at $\unit[401]{nm}$ (\textit{Nichia NDV4B16}) and strontium at $\unit[461]{nm}$ (\textit{Nichia NDB4916}) and $\unit[689]{nm}$ (\textit{Ushio HL69001DG}).
After optimization of the algorithm's parameters, reliable injection locking could be obtained for all tested laser diodes.
It is important to note, that the thermal response of the slave diodes is notably stronger for the blue diodes, such that slower ramps may be necessary.
However, this is dependent on the tuning parameters, and we required maximally $\unit[1]{s}$ for the algorithm to finish without optimizing for time.\\

\section*{Conclusion}

In summary, we have demonstrated a highly reliable and cost-efficient technique to achieve spectrally pure injection locking for a variety of different laser diodes.
Our simple, but general scheme can easily be implemented in a wide range of applications, that require quasi-CW laser light.
Its low-cost implementation makes it especially appealing and allows for scalability of the system.
Even serial systems proof to be stable and need very low maintenance in everyday operation.
This makes our injection locking technique an interesting choice for various applications where narrow linewidth laser light with hundreds of mW output power in several beam delivery paths is required.\\

\begin{acknowledgments}
	We acknowledge funding from the Horizon Europe programme HORIZON-CL4-2022-QUANTUM-02-SGA via the project 101113690 (PASQuanS2.1) and the Federal Ministry of Education and Research Germany (BMBF) via the projects FermiQP (13N15889) and MUNIQC-ATOMS (13N16086).
	We also acknowledge funding from the Deutsche Forschungsgemeinschaft within the research unit FOR5522 (Grant No. 499180199), a Heisenberg professorship to CG and funding from the Alfried Krupp von Bohlen and Halbach foundation.
	The authors would like to thank the Strontium-team of our group (Valerio Amico, Jackson Ang'ong'a, Roberto Franco and Xintong Su) for implementing and testing our findings on $\unit[461]{nm}$ and $\unit[689]{nm}$ laser diodes.
\end{acknowledgments}

\bibliography{sn-bibliography}

\end{document}